\documentclass[aps,reprint,graphicx,longbibliography,superscriptaddress]{revtex4-1}
\usepackage{dcolumn}
\usepackage{bm}
\usepackage{natbib}
\usepackage[usenames,dvipsnames] {color}
\usepackage[table]{xcolor}
\usepackage{graphicx}
\usepackage{epstopdf}
\usepackage{amssymb}
\usepackage{amsmath}
\usepackage{dcolumn}
\usepackage{subfigure}
\usepackage{setspace}
\usepackage[font=singlespacing,font=footnotesize, justification=RaggedRight]{caption}
\usepackage{enumitem}
\setlist{nolistsep}   
\usepackage{hyperref}
\hypersetup{
  colorlinks   = true, 
  urlcolor     = blue, 
  linkcolor    = blue, 
  citecolor   = blue 
}    

\usepackage{blindtext}    
\usepackage{listings}
\usepackage{color} 
\definecolor{mygreen}{RGB}{28,172,0}
\definecolor{mylilas}{RGB}{170,55,241}
\DeclareGraphicsExtensions{.eps,.png,.pdf}

\begin{document}             

\title{Extracting the equation of state of lattice gases from Random Sequential Adsorption simulations by means of the Gibbs adsorption isotherm}
\date{\today}

\author{Shaghayegh Darjani} 
\affiliation{Energy Institute and Department of Chemical Engineering, City College of the City University of New York, New York, NY 10031,
USA.}

\author{Joel Koplik} 
\affiliation{Benjamin Levich Institute and Department of Physics, City College of the City University of New York, NY 10031, USA.}

\author{Vincent Pauchard} 
\email{vpauchard@ccny.cuny.edu.}
\affiliation{Energy Institute and Department of Chemical Engineering, City College of the City University of New York, New York, NY 10031,
USA.}

\begin{abstract}
A novel approach for deriving the equation of state for a 2D lattice gas is proposed, based on arguments similar to those used in the derivation of the Langmuir-Szyszkowski equation of state for localized adsorption. The relationship between surface coverage and excluded area is first extracted from Random Sequential Adsorption simulations incorporating surface diffusion (RSAD). The adsorption isotherm is then obtained using kinetic arguments and the Gibbs equation gives the relation between surface pressure and coverage. Provided surface diffusion is fast enough to ensure internal equilibrium within the monolayer during the RSAD simulations, the resulting equations of state are very close to the most accurate equivalents obtained by cumbersome thermodynamic methods. An internal test of the accuracy of the method is obtained by noting that adsorption RSAD simulations starting from an empty lattice and desorption simulations starting from a full lattice provide convergent upper and lower bounds on the surface pressure.
\end{abstract}
\maketitle

\section{introduction}
Starting from the pioneering work of Langmuir \cite{langmuir1918adsorption}, monolayer models have proved to be extremely useful in explaining physical phenomena such as the adsorption of gases on solid substrates and the decrease in interfacial tension between fluids in the presence of surfactants \cite{mittal2007freundlich,belton1976langmuir,hameed2007adsorption,rane2015_coal_synthetic_asp,rheology_rane_2013,pauchard2014asphaltene,liu2017mixture,bottcher2017saponins,stanimirova2011surface}. Langmuir's model is theoretically restricted to localized adsorption, where adsorbate molecules are smaller than adsorption sites \cite{foo2010insights,bottcher2017saponins}. This strong assumption is violated in many practical cases and in particular for surfactants, which are generally distinctly larger than water molecules. Likewise, the Volmer model \cite{bottcher2017saponins,rosen1989surfactants,kralchevsky1997chemical}, which assumes fully delocalized adsorption (adsorbates much bigger than adsorption sites), is equally inappropriate except, perhaps, for nanoparticles. The intermediate situation, where an adsorbate molecule occupies a few adsorption sites, is better described by Lattice Gas models \cite{baxter,fisher1963lattice,lee1952statistical}, which can readily treat the dynamics of finite-sized objects. These models successfully capture significant experimental features such as phase transitions at high relative coverage and slow adsorption kinetics due to excluded area effects, but have not so far been used in surfactants studies. Among the reasons limiting their practical use is the absence of analytical forms for both the adsorption isotherm and the equation of state and the difficulty in deriving these fundamental equations for each particular adsorbate size and shape case of interest. We can illustrate the difficulty by considering the equation of state side of the problem -- the relationship between surface coverage and surface pressure. \par
Thermodynamically, the surface pressure is given by:
\begin{equation}
\Pi  =  - {\left( {\frac{{\partial {E_H}}}{{\partial A}}} \right)_{N,T}} =  - \frac{1}{{{A_0}}}{\left( {\frac{{\partial {E_H}}}{{\partial M}}}
\right)_{N,T}}
\label{pi1}
\end{equation}
Here, $E_H$  is Helmholtz free energy, $A$ and $A_0$ are the area of the monolayer and the individual area of an adsorption site, respectively, M is the number of sites, T is the absolute temperature and N is the number of adsorbed molecules. The Helmholtz free energy of the monolayer, $E_H$, is itself given by:
\begin{equation}
{E_H} =  - kT\ln (Q)
\label{Helmholtz}
\end{equation}
where $k$ is Boltzmann's constant and $Q$ is the partition function of the monolayer. The determination of surface pressure thus boils down to the determination of the partition function $Q$, the sum over all possible microscopic arrangements of the $N$ adsorbed molecules over $M$ sites, weighted by the energy of the configuration. For simplicity, we consider ideal adsorption, assuming that the energy is independent of the adsorption state. The surface pressure can then be written as:
\begin{equation}
\Pi  = \frac{{kT}}{{{A_0}}}{\left( {\frac{{\partial \ln \mathcal{N} }}{{\partial M}}} \right)_{N,T}}.
\label{pi2}
\end{equation}
where $\mathcal{N}$ is the number of adsorbed configurations. In the Langmuir case of one molecule per site:
\begin{equation}
\mathcal{N}  = \frac{{M!}}{{(M - N)!N!}}
\label{configuration}
\end{equation}
and using Stirling's approximation:
\begin{equation}
\Pi  =  - \frac{{kT}}{{{A_0}}}\ln (1 - \frac{N}{M})
\label{pi}
\end{equation}
The result involves only the fractional coverage $\Theta=N/M$. When an adsorbate molecule covers several adsorption sites, this simple analytical derivation is not possible and the number of possible arrangements
has to be estimated for each fractional coverage value for the actual geometry of the system. A number of methods have been developed over the past six decades, with various degrees of approximation:
\begin{itemize}
\item {The matrix method of Kramers and Wannier
\cite{runnels1966_sqt_tri,runnels1967honeycomb,runnels1970_22square,runnels1970dimer,ree_chestnut1966_sq,orban1968,karayianis1966sqf,bellemans1966phase,temperley1962statistics,nisbet1974first,wood1980vertex,kinzel1981extent,pearce1988classical,guo2002finite},}
\item {The density (or activity) series expansion
\cite{orban1968,gaunt1967tri_lattice,gaunt1965hard,bellemans1967sq_lattice,fisher1963lattice,baxter1980hard,baram1994hard},}
\item {The generalized Bethe
method \cite{cowley1979bethe,burley1961_sq,bellemans1967sq_lattice,runnels1967phase,woodbury1967noncombinatorial,robledo1991hard,hansen2005hard},}
\item{The Rushbrooke and Scoins method \cite{rushbrooke1955ising},}
\item{Monte Carlo
simulation \cite{chesnut1971monte,binder1980phase,meirovitch1983monte,hu1989percolation,haas1989numerical,yamagata1995critical,heringa1996simple,heringa1998cluster,liu2000ordering},}
\item{Fundamental measure theory \cite{lafuente_2003_sq_tri}.}
\end{itemize}
The actual calculations required by these methods are cumbersome and sometimes disagree, particularly at high coverage when lattice gases often exhibit a phase transition to an ordered phase. Furthermore, in most cases the method is ``one-shot€™'' and cannot be systematically improved. \par
An alternate route for calculating the Lattice Gas equations of state is based upon the Gibbs adsorption isotherm \cite{aveyard1973introduction}. If a 2D lattice gas is in equilibrium with a 3D solution of adsorbate molecules, equality of chemical potential throughout the system leads to:
\begin{equation}
d\Pi = kT\,\frac{\Theta}{A_a}\,d\ln C
\label{1}
\end{equation}
where $A_a$ is the interfacial area covered by a single adsorbate molecule and $C$ is the concentration of the 3D solution. Integrating, 
\begin{equation}
\int\limits_0^\Theta  {\frac{\Theta }{C}} \,\frac{{\partial C}}{{\partial \Theta }}\,d\Theta  = \frac{{{A_a}}}{{kT}}\,\Pi 
\label{2}
\end{equation}
From this equation we see that knowledge of the adsorption isotherm, the relationship between $C(\Theta)$, bulk concentration, and fractional coverage, enables one to calculate the equation of state $\Pi(\Theta)$.\par 
The adsorption isotherm, in turn, can be obtained through kinetic arguments. At equilibrium the rates of adsorption and desorption of molecules are equal:
\begin{equation}
K_a\,C(1-\beta(\Theta)) = K_d\,\Theta
\label{3}
\end{equation}
where $K_a$ and $K_d$ are the adsorption and desorption rate constants, respectively, and $\beta (\Theta )$ is the ``blocking function€™", the fraction of the surface area which is excluded from further adsorption by already adsorbed molecules. Solving for $C$ and inserting the result into the integral version of the Gibbs adsorption isotherm yields:
\begin{equation}
\int\limits_0^\Theta  {(1 - \beta (} \Theta ))
\frac{\partial }{{\partial \Theta }}
\left[\frac{\Theta }{1 - \beta (\Theta )}\right]\,d\Theta  
= \frac{{{A_a}}}{{kT}}\,\Pi 
\label{5}
\end{equation}
Then, determining the blocking function is the only information needed to calculate the equation of state.\par
In the Langmuir case, a molecule adsorbed on a site only prevents adsorption on the same site, so that $\beta (\Theta )=\Theta$. If this expression is inserted into Eq.~\ref{5}, we find the same equation of state as determined previously from the partition function (provided the area of the adsorption site and the area of the adsorbates are the same). For lattice gases the blocking function is not known in any simple analytic form, but can be easily extracted from Random Sequential Adsorption (RSA) model simulations.\par
In RSA \cite{evans1993rsa}, objects adsorb onto the open sites of a one or two-dimensional lattice, under specified constraints, using a Monte Carlo type of simulation procedure. The model has been widely used in applications ranging from chemisorption, deposition, layered growth and the car parking problem \cite{wang1993RSA_sq_f, wang1993_sq22, budinski2003ads_des_dif, krapivsky1994reversible_rsa, adamczyk1995kinetics, talbot2000_rsa_review, vscepanovic2013relaxation, budinski2005symmetry, cadilhe2007random, lonvcarevic2007reversible, vscepanovic2016response, evans1993rsa}. In our implementation, molecules or particles are progressively added to an initially empty surface with the only restriction being that overlap is not allowed, an assumption based physically on short range electrostatic repulsion.  As the coverage increases, the free area left for further adsorption decreases, not only because of the area occupied by previously adsorbed molecules but also because vacancies can be too small to allow adsorption without overlap. Without desorption or surface diffusion, adsorption kinetics rapidly slow down and coverage asymptotically approaches the jamming limit, which is the same as random maximum packing if the substrate is not pre-patterned. In the RSAD model, when surface diffusion is introduced in parallel with adsorption, vacancies large enough to adsorb a further particle are both created and destroyed. When diffusion is sufficiently rapid, the size distribution of vacancies no longer depends on the history of adsorption (the positions where adsorbates first arrived on the substrate) but only on the fractional coverage.  In other words, when diffusion is fast enough, the surface layer is at internal equilibrium (even during transient adsorption) and the blocking function can be considered as a state function. From the definition of the adsorption rate used above to define adsorption equilibrium, the blocking function can be extracted from the numerical simulations through the derivative of coverage with respect to the number of attempts:
\begin{equation}
\frac{{\partial N}}{{\partial n}} = 1 - \beta (\Theta )
\label{6}
\end{equation}
where $n$ is the number of attempts ($nA_a/A$=$K_aCt$, is an adimensional time defined via the adsorption rate). This definition will be used in practice by equating the blocking function to the rebuttal rate of adsorption attempts.\par
A practical difficulty in obtaining the blocking function is that because diffusion asymptotes very slowly to its final state in densely packed geometries, simulations can become slow and computationally costly. To address this problem, we will obtain upper and lower bounds using two complementary RSAD simulations: an ``adsorption method€™" which begins from an empty lattice and a new ``desorption method" which begins with a full lattice and progressively decreases coverage (details given in following section). In the first case, the adsorption history has a high randomness in adsorbates relative position: for a slow diffusion the blocking function will be higher than it would be at internal equilibrium. Without diffusion the blocking function would even reach a value of 1 for a fractional coverage corresponding to the random maximum packing. In the second case, to the contrary, the adsorption history has too high an order in adsorbates relative position: for a slow diffusion the blocking function will be lower than it would be at internal equilibrium. Without diffusion the blocking function would just follow that of the Langmuir model. For each coverage, the first and second methods then give upper and lower bounds of the blocking function and by means of integration, equation of state is obtained.\par
An example of actual simulations is given in next section and compared to previously published data. 

\section{Simulation details -- Triangular lattice with nearest neighbor exclusion}
\begin{figure}[!htb]
\centering
\includegraphics[width=0.48\textwidth]{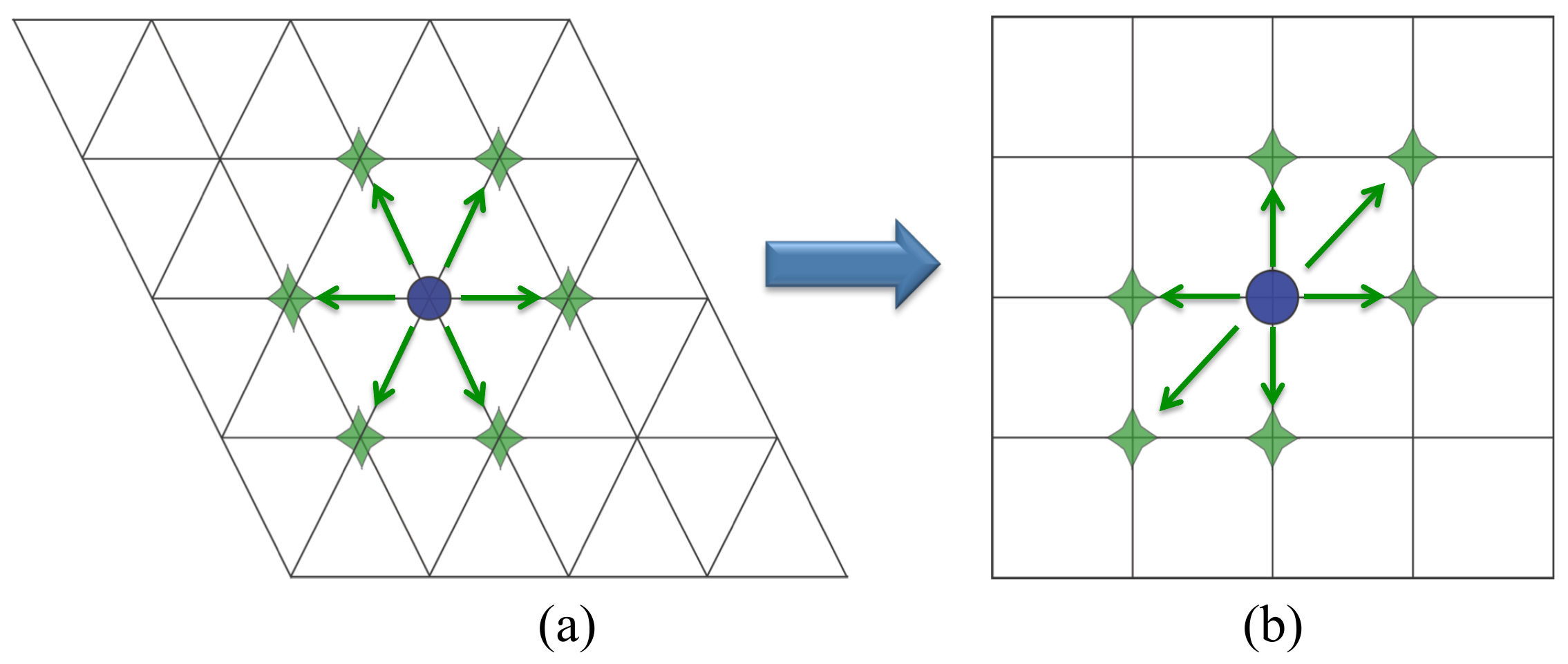}	   
\caption{\footnotesize{(a): triangular lattice with nearest neighbor exclusion (the center of the hexagonal molecule is represented by a circle, the
nearest neighbors by a star). (b): equivalent model in a square geometry. Arrows indicate possible displacement of particles.}}
\label{tri_shape} 
\end{figure}

The RSA model on a triangular lattice with nearest neighbor exclusion case has been extensively studied in the literature using a variety of methods \cite{burley1960lattice,burley1965first,baxter,runnels1966_sqt_tri,gaunt1967tri_lattice,orban1968,lafuente_2003_sq_tri,gaunt1967tri_lattice,cowley1979bethe,chesnut1971monte}. The model involves the adsorption of hexagonal molecules covering three adsorption sites, in the following sense. If the center of a hexagon occupies a node of the lattice, the center of another hexagon cannot occupy any of the six immediate neighboring nodes. For each adsorbate the center node is counted in full but each of the six neighboring nodes can be shared with 2 other adsorbates. The average occupancy is then 1+6/3=3, or down to 3 nodes covered per adsorbate.  For computational convenience a triangular lattice of dimensions d$\times$d is converted into a square lattice of same dimensions as shown in Figure~\ref{tri_shape}. To limit finite size effects, periodic boundary conditions are used in the simulations.
For the first method the lattice is initially empty and hexagons are progressively added. For each adsorption attempt, a random position (x,y) is selected representing the center of the hexagon. If position (x,y) does not infringe the non-overlap condition, adsorption is accepted. Otherwise it is rebutted. Diffusion is introduced sequentially with a predefined ratio D between the number of diffusion attempts and the number of adsorption attempts: for D=1 each adsorption attempt is followed by a diffusion attempt, for D=5 each adsorption attempt is followed by 5 diffusion attempts, etc. For each diffusion attempt, a previously adsorbed hexagon is randomly selected. A direction for the displacement of the hexagon is also randomly selected among 6 possibilities: 2 along each of the principal directions of the lattice, 2 along each of its diagonals. If moving the center of the hexagon to the next node along this direction does not infringe the non-overlap condition, diffusion is accepted. Otherwise it is rejected. \par
For the second method the lattice is initially full. For each simulation step, 2 hexagons are first randomly selected and removed. Then one adsorption attempt and D diffusion attempts are performed following the same procedure as for the first method. The choice of the sequence (2 desorption events followed by 1 adsorption) is arbitrary but answers the need at each time step to decrease coverage and add at least one molecule to calculate the blocking function.\par
For both methods, the blocking function is extracted from the success rate of adsorption attempts. To reduce the noise arising from the numerical calculation of the derivative of the coverage, 500 runs are performed and averaged for each simulation condition. The blocking function is then fitted with a polynomial function and used to generate the adsorption isotherm. The latter is inserted into the Gibbs adsorption isotherm equation (\ref{5}) to obtain the equation of state. Note that in this model the potential energy is effectively infinite for overlaps, due to the repulsive interaction which restricts the occupancy of neighbors, and is zero otherwise, so the system can be considered as athermal \cite{fernandes2007monte}. \par
\section{Results}
\begin{figure}[!htb]
\centering
\includegraphics[width=0.48\textwidth]{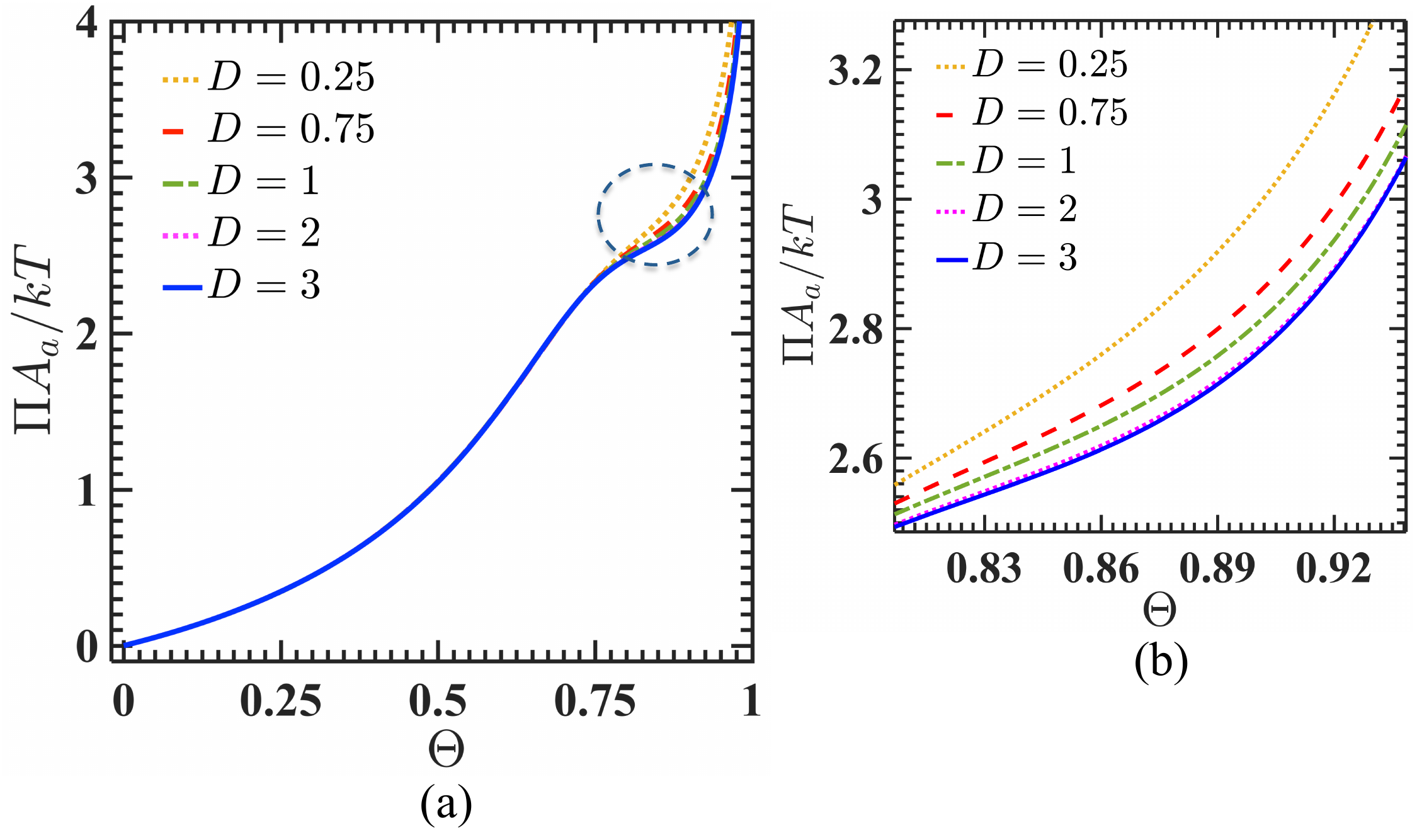}	   
\caption{\footnotesize{(a): Triangular lattice surface pressure by the adsorption method: effect of diffusion on the equation of state for lattice size of $d=99$. The parameter {\em D} is the relative frequency of diffusion to insertion attempts, (b): the inset expands the phase transition region where sensitivity to diffusion appears.}}
\label{tri_fm_dif} 
\end{figure}

The effect of surface diffusion on the equation of state (EOS) using the adsorption method is studied in Figure~\ref{tri_fm_dif} for lattice size of $d=99$. At low surface coverage when the system is dilute, molecules can easily find open positions in the surface with no difficultly due to blocking, so diffusion has little effect and all curves overlap initially. At high surface coverage, in contrast, rearrangement may be needed to open gaps on the lattice. Increasing the surface diffusion rate leads to reorganization and provides space for new incoming particles. When the surface diffusion is large enough the system reaches equilibrium. Note that the practical effect of increasing the diffusion rate is to lower the surface pressure.\par
\begin{figure}[!htb]
\centering
\includegraphics[width=0.48\textwidth]{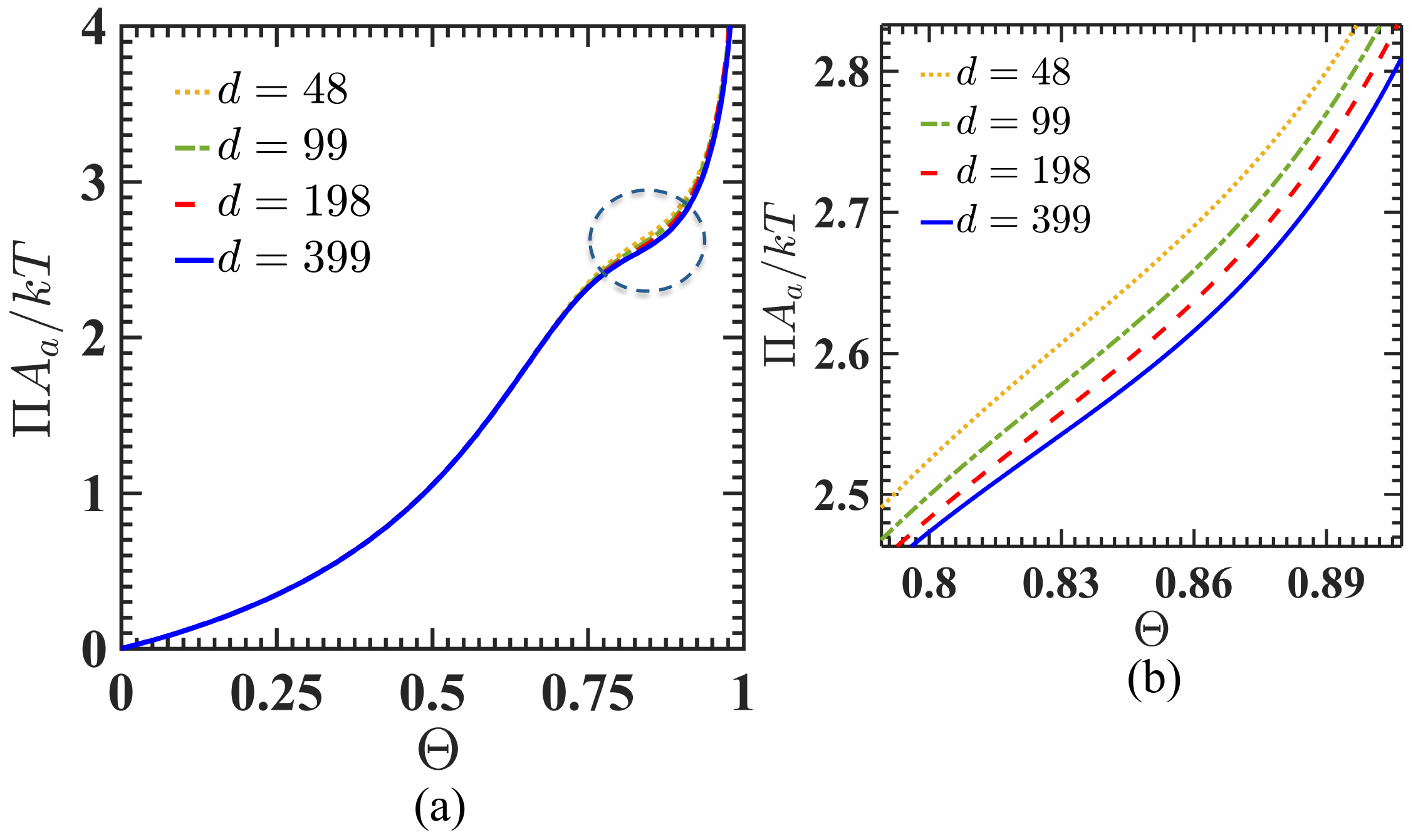}	   
\caption{\footnotesize{(a): Triangular lattice surface pressure by the adsorption method: effect of lattice size on equation of state for surface diffusion of $D=1$, (b): the inset expands the phase transition region where sensitivity to lattice size appears.}}
\label{tri_fm_size} 
\end{figure}

\begin{figure}[!htb]
\centering
\includegraphics[width=0.48\textwidth]{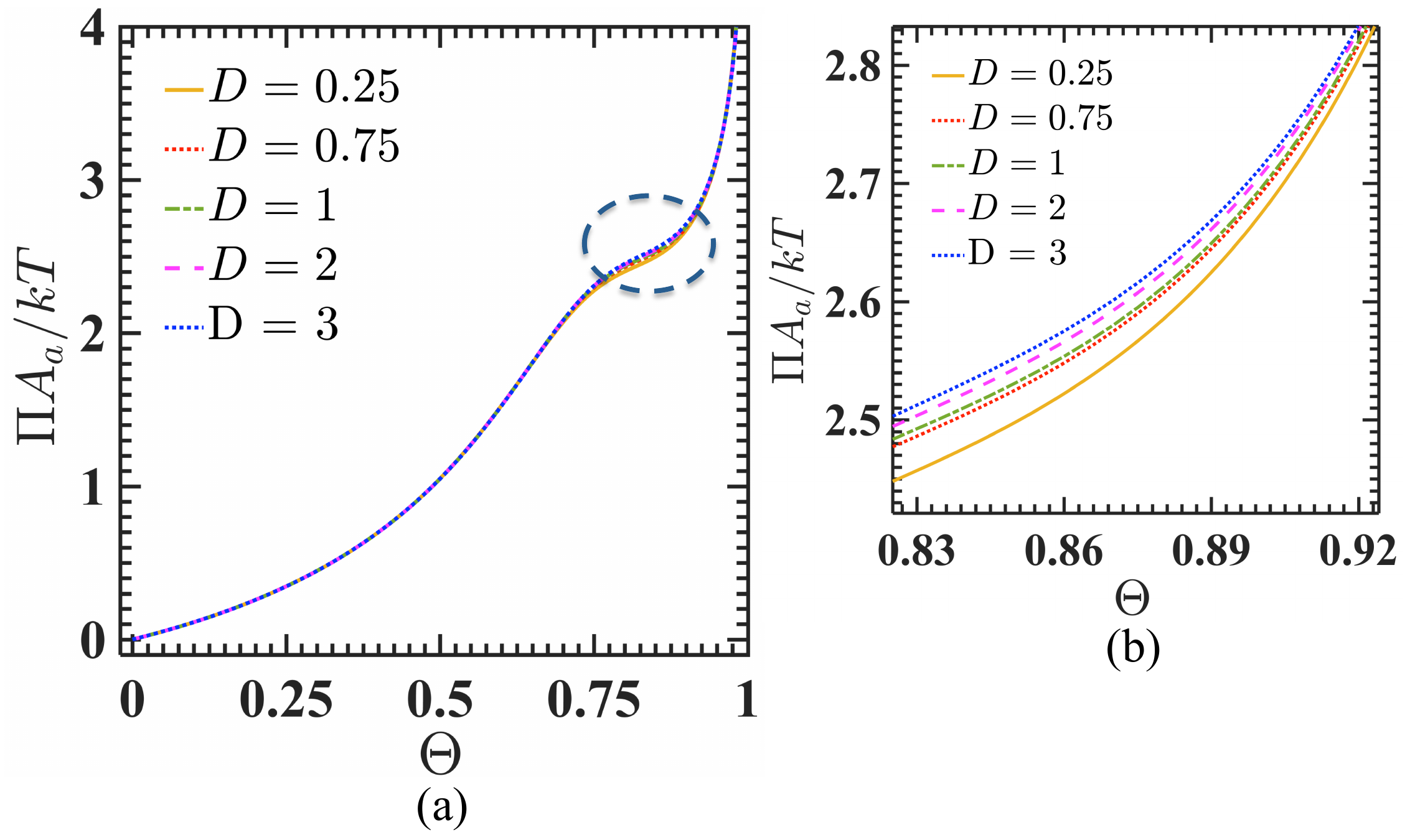}	   
\caption{\footnotesize{(a): Triangular lattice surface pressure by the desorption method: effect of diffusion on the equation of state for lattice size of $d=99$, (b): the inset expands the phase transition region where sensitivity to diffusion appears.}}
\label{tri_sm_dif} 
\end{figure}

\begin{figure}[!htb]
\centering
\includegraphics[width=0.48\textwidth]{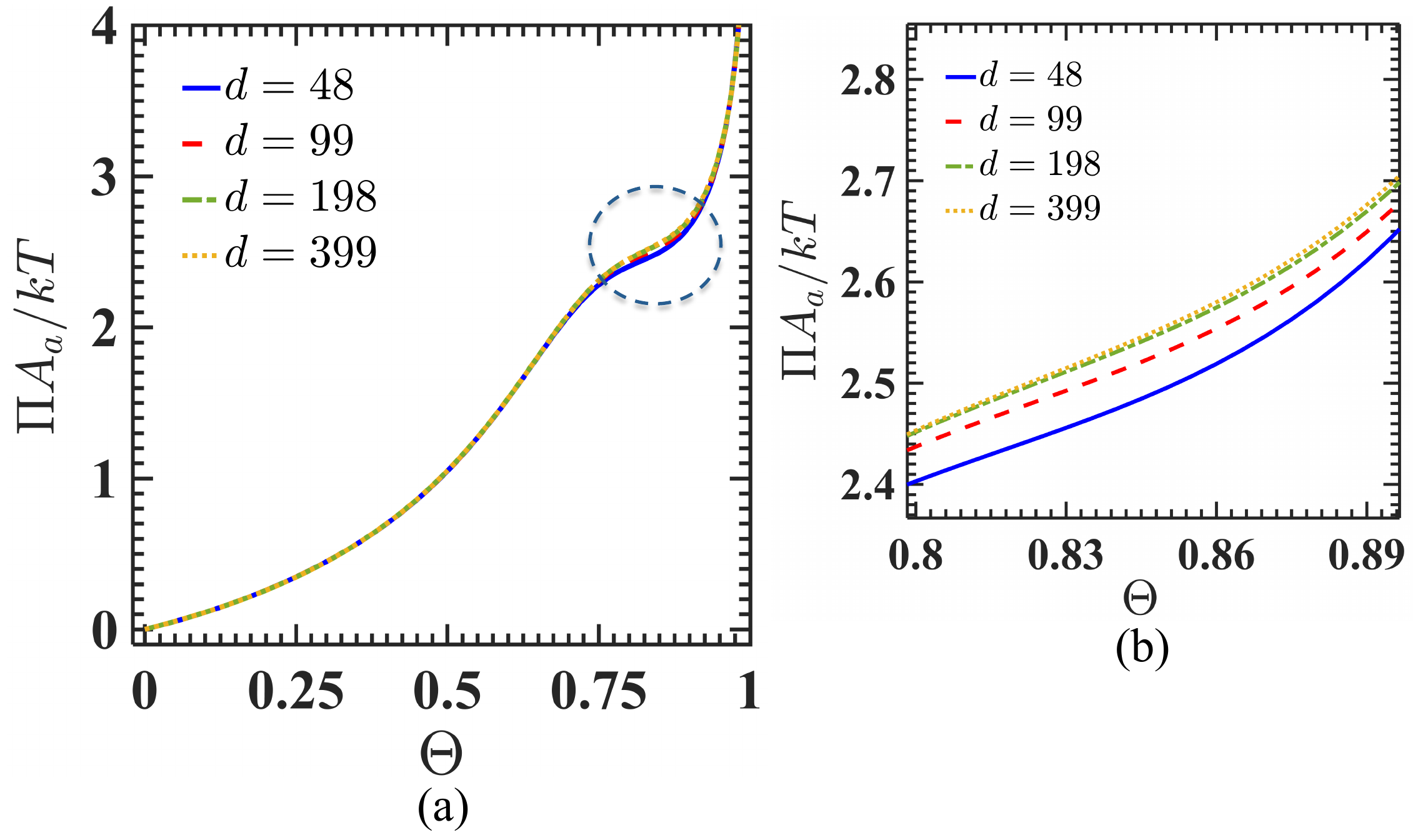}	   
\caption{\footnotesize{(a): Triangular lattice surface pressure by the desorption method: effect of lattice size on the equation of state for surface diffusion of $D=1$, (b):the inset expands the phase transition region where sensitivity to lattice size appears.}}
\label{tri_sm_size} 
\end{figure}

A key issue in lattice simulations is the effect of finite system size on the equation of state. Figure~\ref{tri_fm_size} presents the variation in EOS for triangular lattices of sizes 48 to 399 for surface diffusion of $D=1$: the surface pressure at first decreases with increasing lattice size but eventually converges.
The analogous results for the desorption method are shown in Figures~\ref{tri_sm_dif}-\ref{tri_sm_size}. These results agree with those of the adsorption method at low surface coverage, when the system is independent of surface diffusion. However, in the vicinity of the phase transition, the trends are opposite, meaning that higher surface diffusion and lattice size give higher surface pressure. Eventually, all curves obtained using the desorption method overlap for sufficiently large diffusion and lattice size.\par
\begin{figure}[!htb]
\centering
\includegraphics[width=0.48\textwidth]{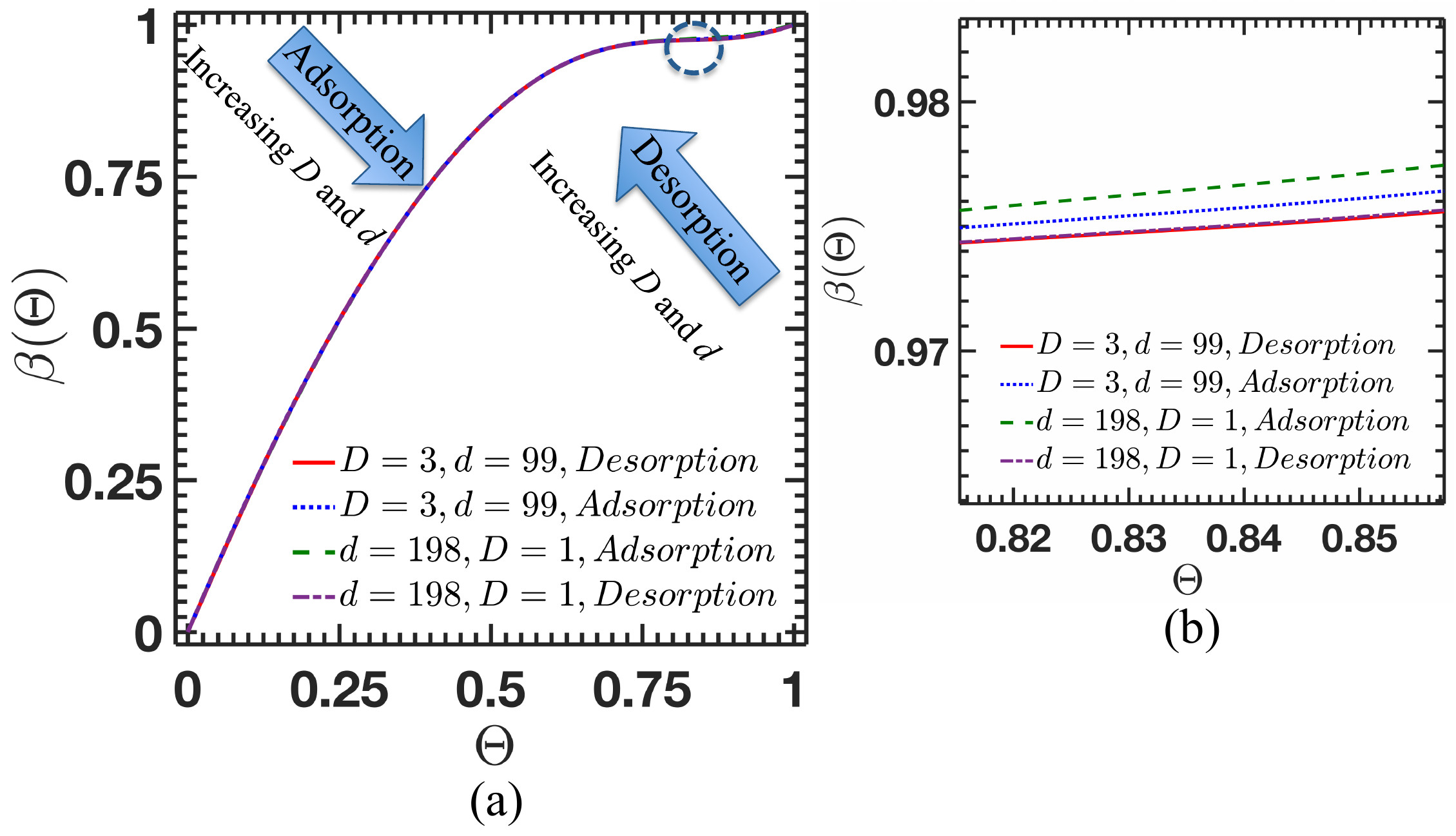}	   
\caption{\footnotesize{(a): Blocking Function, (b): the inset shows a magnified view of the region bounded by dark blue dashed circle.}}
\label{blocking_function} 
\end{figure}

The blocking function provides a clear illustration of the convergence of the adsorption and desorption methods, as shown in Figure~\ref{blocking_function}. We expect that the equilibrium equation of state should be bracketed by the results of the two methods, and to pursue this idea, it is useful to compare our results with the analytic calculations of Baxter \cite{baxter} and Runnel \cite{runnels1966_sqt_tri}.\par
\begin{figure}[!htb]
\centering
\includegraphics[width=0.48\textwidth]{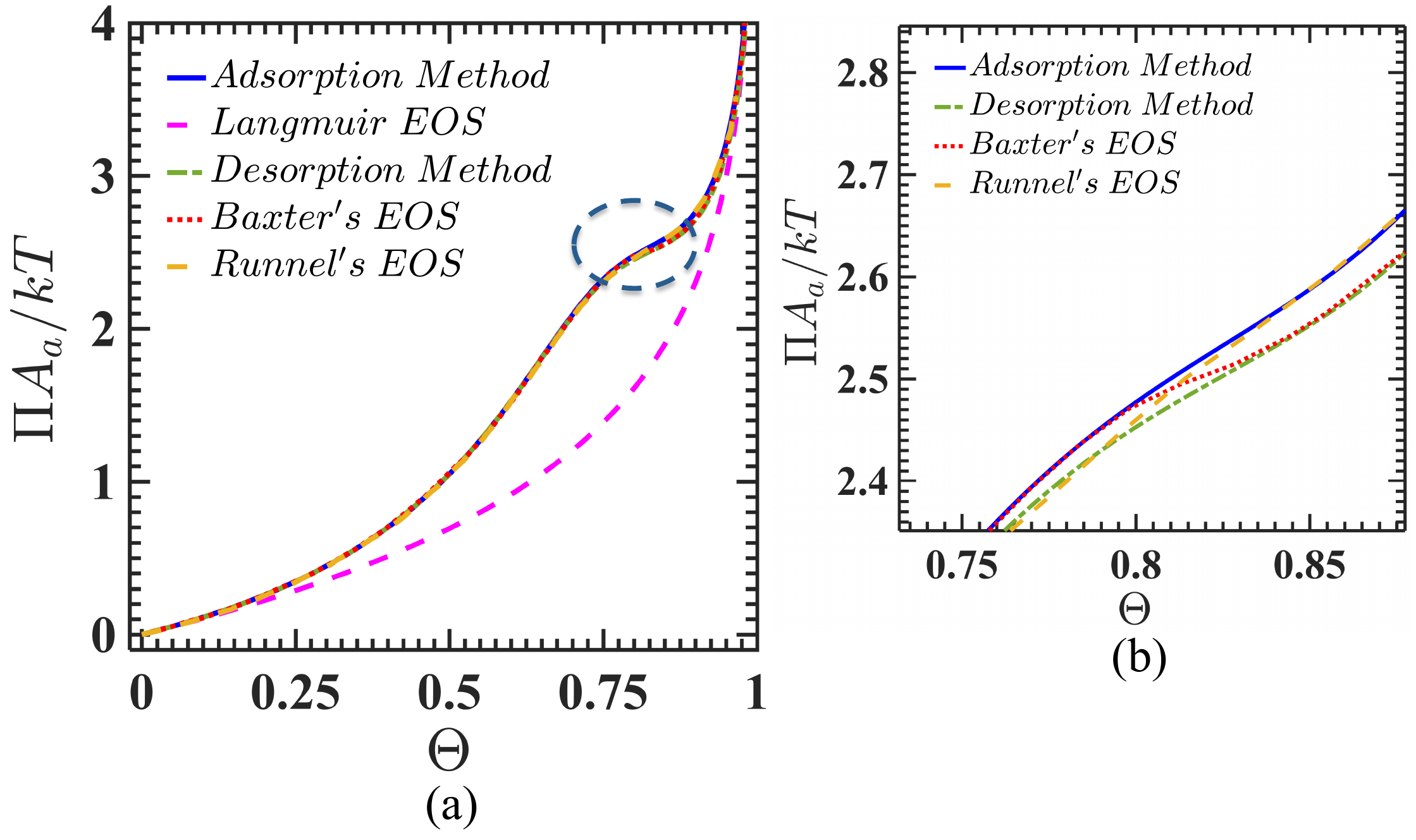}
\caption{\footnotesize{(a): Comparison between our EOS where d=99 and D=4 with those of Baxter \cite{baxter} and Runnels \cite{runnels1966_sqt_tri}) for hard core molecules on a triangular lattice, (b): the inset shows a magnified view of the region bounded by dark blue dashed circle.}}
\label{EOS} 
\end{figure}

Baxter \cite{baxter} studied the partition function of the eight vertex model, which has the same transfer matrix as the hard hexagon model, in the limit of infinite lattice size. He found that at critical activity ($z_c=11.0902$), the partition function per site depends only on this parameter and the resulting critical density and pressure were 0.829 and 2.5175, respectively. Runnels \cite{runnels1966_sqt_tri} used an Exact Finite Matrix method, based on a sequence of exact solutions for lattices of infinite length and increasing finite width. The results were that, far from the transition zone, convergence occurs rapidly; while in the transition region, thermodynamic properties such as density and pressure are only functions of lattice width which can be extrapolated to infinite width, giving the critical density and surface pressure as 0.837 and 2.529, respectively. The EOS results using various methods are compared in Figure~\ref{EOS} where $d$ is equal to 99 and $D=4$.
The results of Baxter and Runnels overlap with our EOS data at low coverage, but in the vicinity of the phase transition a slight difference is observed. Baxter's EOS follows the adsorption method at lower $\Theta$ and the desorption method at higher $\Theta$, while Runnels' EOS has the opposite behavior. Eventually, however, all of the calculations overlap the Langmuir EOS around the maximum packing coverage.\par
\begin{figure}[!htb]
\centering
\includegraphics[width=0.25\textwidth]{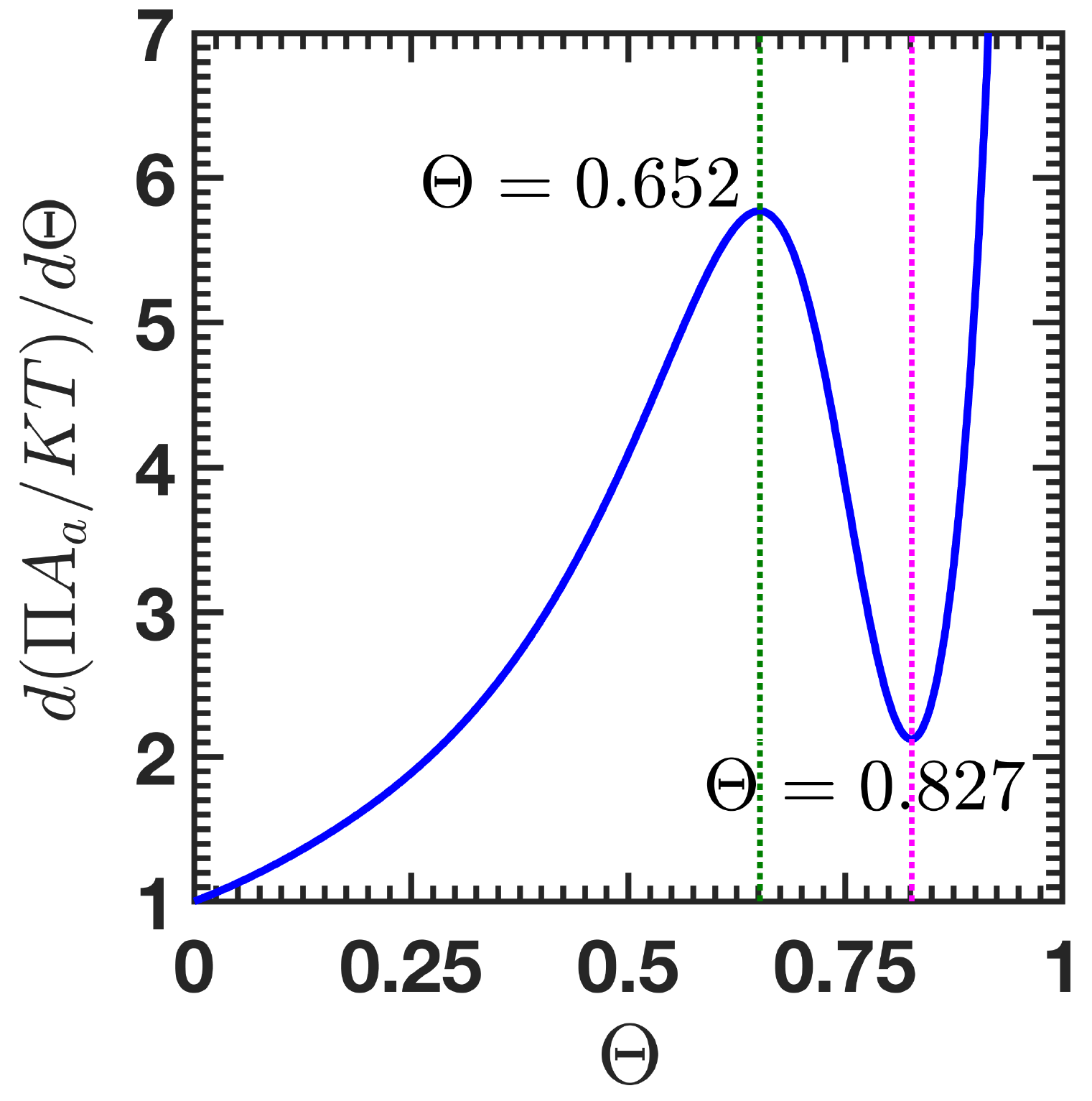}	   
\caption{\footnotesize{Analysis of phase transition region where $d=99$ and $D=4$.}}
\label{phase_transition} 
\end{figure}

 To further verify our methods, we can examine the phase transition zone in more detail. The derivative of the surface pressure with respect to surface coverage can be computed from the simulation data, with the result depicted in Figure~\ref{phase_transition}. A second order phase transition is obtained for this system, a hard core molecule with first neighbor exclusion on a triangular lattice with $d=99$ and $D=4$, where the inset of deviation from liquid regime starts at around  $\Theta=0.652$ and system solidifies at around $\Theta=0.827$. These values are in excellent agreement with those of Baxter \cite{baxter}. 
\section{Conclusion} 
Random sequential adsorption with surface diffusion was used to find the relationship between the blocking function and surface coverage. The results show that RSAD can be used as an equilibrium model provided that surface diffusion is rapid enough to ensure that internal equilibrium is reached before each deposition. At internal equilibrium, the blocking function can be considered as a state function. The equilibrium blocking function then can be used to produce both adsorption isotherm and equation of state by coupling kinetic arguments and the Gibbs equation. We thus obtain a unique master curve representing isothermal equilibrium states where the EOS is independent of the initial configuration and the saturation path.\par
We employ two complementary methods, which start from either empty or saturated lattice configurations and are expected to bracket the correct equilibrium equation of state. Our methods compared favorably with two statistical mechanic calculations with the same geometry. The results of Baxter \cite{baxter} and Runnel \cite{runnels1966_sqt_tri} overlap with both of our methods up to the fractional surface coverage where the deviation from the liquid regime begins, near $\Theta=0.65$. In the vicinity of the phase transition, Baxter's EOS \cite{baxter} overlaps with the adsorption method at the lower bound of this regime and the desorption method at the upper bound, while Runnel's \cite{runnels1966_sqt_tri} EOS shows the opposite behavior. A second order phase transition is found for both approaches, where solidification occurs near $\Theta=0.83$, which is in excellent agreement with Baxter value \cite{baxter}.\par

\section*{ACKNOWLEDGMENTS}
We thank Professor Sanjoy Banerjee for thoughtful discussions and Nelya Akhmetkhanova for proofreading our manuscript.         


%
\end{document}